\documentclass[aip,superscriptaddress,twocolumn,reprint]{revtex4-1}

\usepackage[applemac]{inputenc}
\usepackage{amssymb}
\usepackage{natbib}
\usepackage{amsmath}
\usepackage{amsfonts}
\usepackage{graphicx}
\usepackage{subfigure}
\usepackage{mathrsfs}
\usepackage{dcolumn} 
\usepackage{bm}
\usepackage{color}
\usepackage{cancel}
\usepackage{mathbbol}
\usepackage{epsfig}
\usepackage{units}
\usepackage{esint}
\usepackage{soul} 
\usepackage{url}
\usepackage{afterpage}
\usepackage{hyperref}
\usepackage{physics}

\def\Tr{\mbox{Tr}} 
\newcommand{\av}[1]{\langle#1\rangle}

\newcommand{\me}[1]{\left\langle#1\right\rangle}

\begin{document}

\title{Quantum thermodynamic methods to purify a qubit on a quantum processing unit} 
 
\author{Andrea Solfanelli}
\email{asolfane@sissa.it}
\affiliation{SISSA, via Bonomea 265, I-34136 Trieste, Italy}
\affiliation{INFN, Sezione di Trieste, I-34151 Trieste, Italy}

\author{Alessandro Santini}
\email{asantini@sissa.it}
\affiliation{SISSA, via Bonomea 265, I-34136 Trieste, Italy}

\author{Michele Campisi}
\email{michele.campisi@nano.cnr.it}
\affiliation{NEST, Istituto Nanoscienze-CNR and Scuola Normale Superiore, I-56127 Pisa, Italy}
\affiliation{Department of Physics and Astronomy, University of Florence, I-50019, Sesto Fiorentino (FI), Italy}

\date{\today} 
 
\begin{abstract} 
We report on a quantum thermodynamic method to purify a qubit on a quantum processing unit (QPU) equipped with (nearly) identical qubits. Our starting point is a three qubit design that emulates the well known two qubit swap engine. Similar to standard fridges, the method would allow to cool down a qubit at the expense of heating two other qubits.
A minimal modification thereof leads to a more practical three qubit design that allows for enhanced refrigeration tasks, such as increasing the purity of one qubit at the expense of decreasing the purity of the other two. The method is based on the application of properly designed quantum circuits, and can therefore be run on any gate model quantum computer. We implement it on a publicly available superconducting qubit based QPU, and observe a purification capability down to $200 $ mK. We identify gate noise as the main obstacle towards practical application for quantum computing.
 \end{abstract}
 
 
\maketitle

\section{Introduction}

Quantum computing technology is currently developing at a very fast pace. The main obstacle towards scaling up the number of qubits on the Quantum Processing Unit (QPU) is noise:\cite{Preskill18QUANTUM2} QPU are still subject to a number of noise sources that make them, at the current stage of development, still prone to large error. Noise may affect a quantum computation at each stage thereof, from initial qubit state preparation, to gate application, to read out and storage. Here we focus on the preparation. The starting point of any quantum algorithm, a so called quantum circuit, is a tensor product of the ground states of all qubits on the QPU that participate in the computation. From a thermodynamical perspective that is a zero temperature state. The third law of thermodynamics actually forbids its achievement: such a state can only be achieved to some degree of approximation.\cite{Callen60Book} That is, the unavoidable starting point of any quantum circuit is a state of some finite  (no matter how small) temperature, rather than an ideal pure quantum state. Then, a question of crucial technological relevance is how to achieve smaller and smaller temperature of the initial preparation. The most direct way of addressing this problem is to control and reduce to a minimum all sources of noise that may affect the preparation. 

Here we propose to adopt an alternative thermodynamic approach instead. As we learn from thermodynamics a refrigerator is a machine that takes heat away from a cold body to heat up a hotter one by consuming some power coming from an external energy source.\cite{Fermi56Book} Thus, our idea is to do the same on a QPU, where one qubit would be cooled down at the cost of heating up another qubit (or more qubits as we shall see below), while some energy is spent to make that happen. That energy comes, as we shall see below, from application of a properly designed entangling gate on the set of involved qubits.

\section{Refrigeration Method}

 Our quantum refrigeration scheme is a modification of the so called quantum SWAP engine.\cite{Lloyd97PRA56, Quan07PRE76, Allahverdyan08PRE77,Campisi15NJP17,Timpanaro19PRL123,Uzdin14NJP16} A quantum SWAP engine is composed of two qubits, a hot qubit (labelled as qubit H from now on) being at temperature $T_H$ and a cold one (labelled as qubit C) being at temperature $T_C<T_H$.  As reported previously,\cite{Campisi15NJP17,Buffoni19PRL122} application of the SWAP unitary to the two qubits results in the cold qubit getting to a colder temperature $T'_C<T_C$ and the hot one to a hotter temperature, provided the ratio of the two qubits resonant frequencies $\omega_C/\omega_H$, is smaller than the ratio of their initial temperatures $T_C/T_H$. Besides, among all the unitaries, the SWAP is the one that achieves the highest cooling coefficient of performance (COP), reading $\eta = (\omega_H/\omega_C-1)^{-1} $ . 

The main difficulty that one encounters when trying to implement this simple scheme on current QPUs, is that they are engineered to have ideally identical qubits. If that is the case then the SWAP engine described above would not work: for $\omega_C=\omega_H$ the condition $\omega_C/\omega_H< T_C/T_H$, implies $T_C > T_H$, which contradicts that label $C$ denotes the colder qubit.  
We have evidenced this unfortunate situation with a previous set of experiments performed on an IBM QPU.\cite{Solfanelli21PRXQ2} In order for the cooling mechanism to work, a necessary condition is:
\begin{align}
\omega_H > \omega_C\, .
\end{align}
The larger is $\omega_H$ as compared to $\omega_C$, the more robust will be the cooling operation,\cite{Solfanelli21PRXQ2} while the coefficient of performance $\eta$, will decrease.

The question is then whether one can modify the SWAP engine design, in order to implement a working cooling mechanism on a  QPU with identical qubits with non-tunable resonant frequency.  
One simple way to achieve that is to combine two or more qubits together to form a compound multi-level system having a larger resonant frequency. 

Our solution is then to replace the hot qubit with a compound system made of two qubits, and focus only on its ground and most excited states, $\ket{00}_H$ and $\ket{11}_H$, respectively. Those two states will play the role of a qubit with a doubled resonant frequency. That would make the ``hot'' resonance $\omega_H$ be twice the ``cold'' resonance $\omega_C$ and opens for the possibility of refrigerating the cold qubit.  
\begin{figure}
	\includegraphics[width=\linewidth]{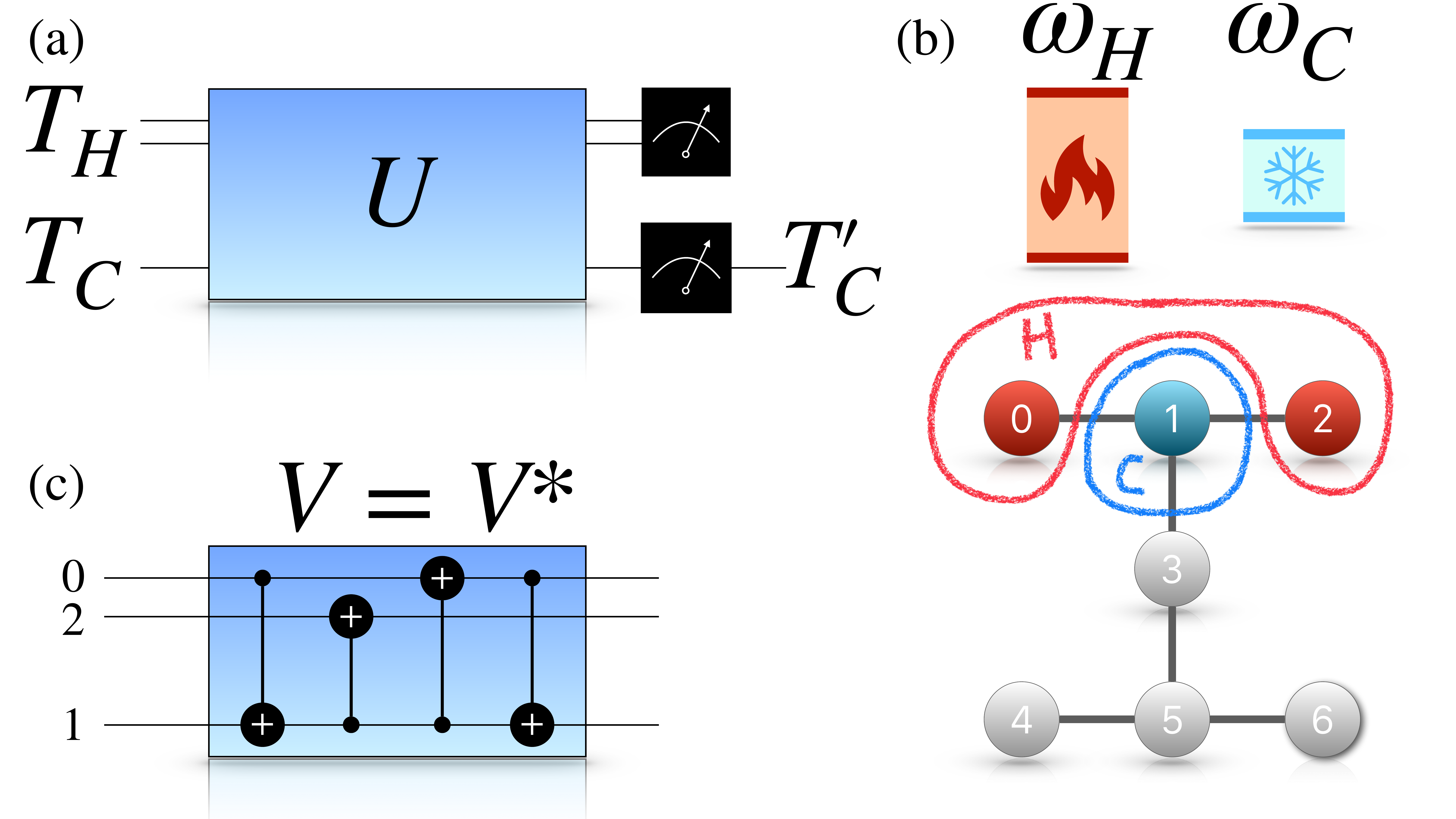}
	\caption{
Panel a): quantum circuit representation of the cooling method. The top two horizontal lines denote the hot subsystem 	compound, initially at temperature $T_H$, the bottom horizontal line represents the cold qubit, initially at temperature $T_C<T_H$. Panel b): topology of the QPU used in the experiments. Qubits 0,2 form the hot subsystem presenting a resonance $\omega_H=\omega_0+\omega_2$. Qubit 1 is the cold qubit that needs to be cooled further.
Panel c): quantum circuit implementation of the unitary gate $U$ with the choice $V=V^*$, on the QPU.
	 }
	\label{fig:figure1}
\end{figure}

\begin{figure*}
	\includegraphics[width=\linewidth]{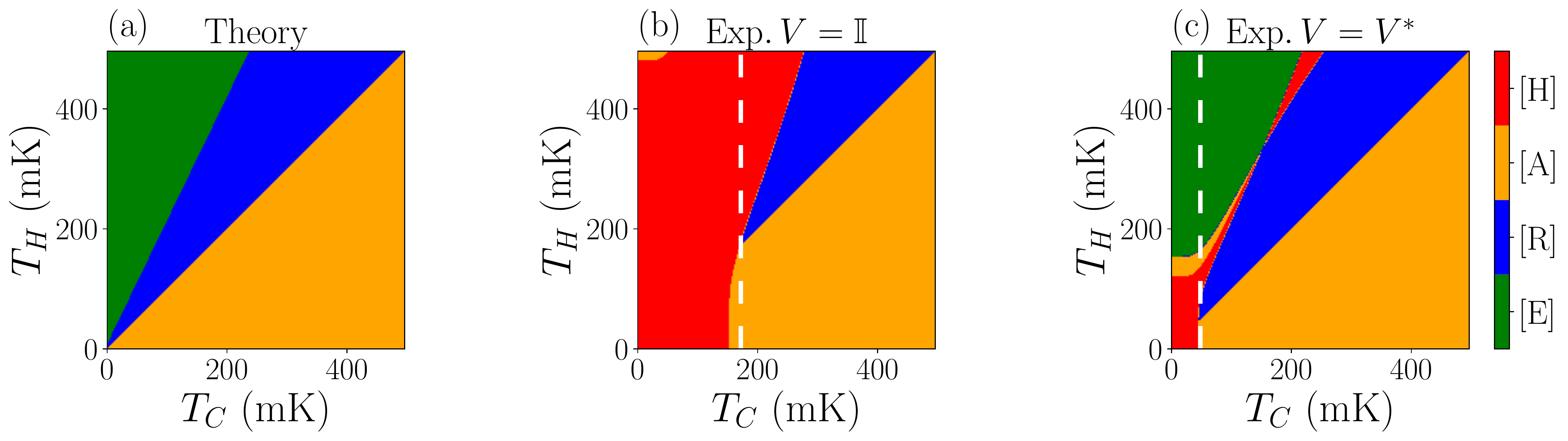}
	\caption{``Phase diagrams'' of the quantum heat engine implementing our refrigeration method. Panel a): Theory. Panel b): Experiment with the choice $V=\mathbb{I}$. Panel c): Experiment with the choice $V=V^*$.}
	\label{fig:figure2}
\end{figure*}

We assume that the qubit to be cooled is initially at some temperature $k T_C=1/\beta_C$ ($k$ is Boltzmann's constant),
and that the states $\ket{00}_H$ and $\ket{11}_H$ are populated according to a Gibbs distribution 
of temperature $k T_H=1/\beta_H$, so that the total system is initially described by the density operator
\begin{align}
\rho = \frac{e^{-\beta_H H_H-\beta_C H_C}}{Z_H Z_C}
\label{eq:prep}
\end{align}
where 
\begin{align}
H_H = -\frac{\hbar\omega_H}{2}\left(\dyad{00}_H-\dyad{11}_H\right) \otimes \mathbb{I}_C 
\label{eq:HH}
\end{align}
with $\omega_H$ being the sum of the resonant frequencies of the two qubits composing system $H$;
and 
\begin{align}
H_C = - \mathbb{I}_H \otimes \frac{\hbar\omega_C}{2}\left(\dyad{0}_C-\dyad{1}_C\right)
\label{eq:HC}
\end{align}
being the cold qubit Hamiltonian ($\hbar$ is the reduced Planck constant). The symbol $Z_r$ stands for the according partition function $Z_r=\Tr_r\,  e^{-\beta H_r}$. $\Tr_r $ and $ \mathbb{I}_r$ denote trace operation and the identity operator in the $r$ subsystem Hilbert space, respectively.
In the following we adopt the notation $\ket{ij,k}$, with $i,j,k=0,1$ to denote the energy eigen-basis of the compound 3 qubit system, with the first two indices referring to system H, and the third index referring to system C.

Given the initial state \ref{eq:prep}, the method consists in applying a unitary operation $U$ that maps $\ket{ii,j}$ onto $\ket{jj,i}$. One such gate generally reads
\begin{align}
U=\left(\begin{array}{c|c} W &  \\\hline  & V \end{array}\right)
\label{eq:U}
\end{align}
where  $V$ is a generic unitary acting on the subspace spanned by 
$\{
\ket{01,0}
\ket{01,1}
\ket{10,0}
\ket{10,1}
\}$
and $W$ is the relevant swap operation occurring in the subspace spanned by $\{
\ket{00,0}
\ket{00,1}
\ket{11,0}
\ket{11,1}
\}$. Its matrix representation in that basis reads
\begin{align}
W=\left(\begin{array}{c|cc|c}e^{i \phi_1} &  &  &  \\\hline  &  & e^{i \phi_2} &  \\  & e^{i \phi_3} &  &  \\\hline  &  &   & e^{i \phi_4}\end{array}\right).
\label{eq:W}
\end{align}
with $\phi_i$ arbitrary phases.

The specific form of $V$ does not have any impact on the thermodynamics of the device. This is because in the preparation of Eq. (\ref{eq:prep}), the states 
$\{
\ket{01,0}
\ket{01,1}
\ket{10,0}
\ket{10,1}
\}$ are not populated, hence any dynamics occurring in the space they span is immaterial from the energetic point of view.
As we shall see below, however, the choice of $V$ may have a great impact from the practical point of view.
The quantum circuit representation of the method is sketched in Fig. \ref{fig:figure1}a)

\subsection{Results}

We have implemented the method on IBM \textit{ibmq\_jakarta} QPU with two different choices of $V$. Its topology is depicted in Fig \ref{fig:figure1}b). Qubit $1$ is the cold qubit that needs to be refrigerated. Qubits $0,2$ form the $H$ system. Their resonant frequencies were $\omega_{0} \approx 5.24\ \mathrm{GHz}$, $\omega_{1} \approx 5.01 \ \mathrm{GHz}$ and $\omega_{2}\approx 5.11\  \mathrm{GHz}$, hence $\omega_C \approx 5.01 \ \mathrm{GHz}, \omega_H  \approx 10.35 \ \mathrm{GHz}$. The methods used to obtain the experimental data are described in detail below.

Figure \ref{fig:figure2}a) shows the theoretical ``phase diagram'' in the $T_H,T_C$ plane showing which thermodynamic operation mode is expected. We recall that, based on general quantum mechanical arguments, only 3 operations modes are possible besides Refrigeration [R].\cite{Solfanelli20PRB101} They are: Heat Engine $[E]$, when heat is transferred from the hot to the cold subsystem while energy is output in the form of work; Thermal Accelerator $[A]$, when heat is transferred from the hot to the cold subsystem while work is spent; Heater $[H]$, when both subsystems receive energy from the work source.
Note the extended connected blue region indicating that refrigeration can in principle be robustly implemented.

Our first choice of $V$ was $V=\mathbb{I}$
with $\mathbb{I}$ denoting the identity operator on the Hilbert space spanned by $\{
\ket{01,0}
\ket{01,1}
\ket{10,0}
\ket{10,1}
\}$. 
Figure \ref{fig:figure2}b) shows the according experimental ``phase diagram'' in the $T_H,T_C$ plane.
Note that in comparison with the theoretical expectation, presenting no region [H]  of heating for both subsystems, a large portion of the phase diagram is in fact taken by this region, especially at low temperatures. A blue [R] region where refrigeration occurred exists, but it is very much shrunk as compared to theory. In particular no refrigeration was observed below temperature $T_C=172\, \mathrm{mK} $ of the cold qubit (dashed vertical line).  
These effects were mostly due to the noise affecting the gate $U$. We note, in fact, that in our experiments the gate $U$ was decomposed and implemented by the IBM compiler as a sequence of more than 180 elementary gates.\footnote{Elementary gates are $R_z(\theta)$, $\sigma_x$, $\sqrt{\sigma_x}$, CNOT, namely single qubit rotation of arbitrary angle $\theta$ around the $z$ axis, the single qubit $\sigma_x$ operator, its square root, and the entangling controlled-NOT operator among two connected qubits.}
Counting that each elementary gate comes with its load of noise, no matter how small, the high number thereof resulted in a good amount of noise, which greatly affected the functioning of the device.

In order to mitigate this problem (and to confirm that gate noise was indeed the source of the detrimental effects) we repeated the experiment with the choice of $V$ being the unitary that maps $\ket{01,1}$ onto $\ket{10,0}$ (and vice-versa), while leaving the states $\ket{01,0}, \ket{10,1}$ unaltered. In the following we shall refer to this choice as $V^*$. Its matrix representation reads, in the basis $\{
\ket{01,0}
\ket{01,1}
\ket{10,0}
\ket{10,1}
\}$,
exactly as $W$ reads in the basis $\{
\ket{00,0}
\ket{00,1}
\ket{11,0}
\ket{11,1}
\}$, i.e., the matrix in Eq. (\ref{eq:W}).

At variance with the $V=\mathbb{I}$ case, with $V=V^*$ the global unitary $U$ was implemented with only $4$ CNOTs as shown in Fig. \ref{fig:figure1}c). 
 
Figure \ref{fig:figure2}c) shows the experimental ``phase diagram'' in the $T_H,T_C$ plane obtained with the choice of $V=V^*$.
Note how this choice has resulted, in comparison with the choice $V=\mathbb{I}$, in a shrinking of the heating region [H] (in red), while the refrigeration region [R] (in blue) has enlarged, thus realising a more robust cooling  operation. Most remarkably, the [R] region now extends down to $T_C= 52\,\mathrm{mK} $ meaning that the improved implementation allows to cool a qubit to lower temperature, as compared to the more noisy case. This clearly indicates that decreasing the gate noise further will lead to even lower cooling temperature, and better performance.

\begin{figure}[h!]
	\includegraphics[width=.75\linewidth]{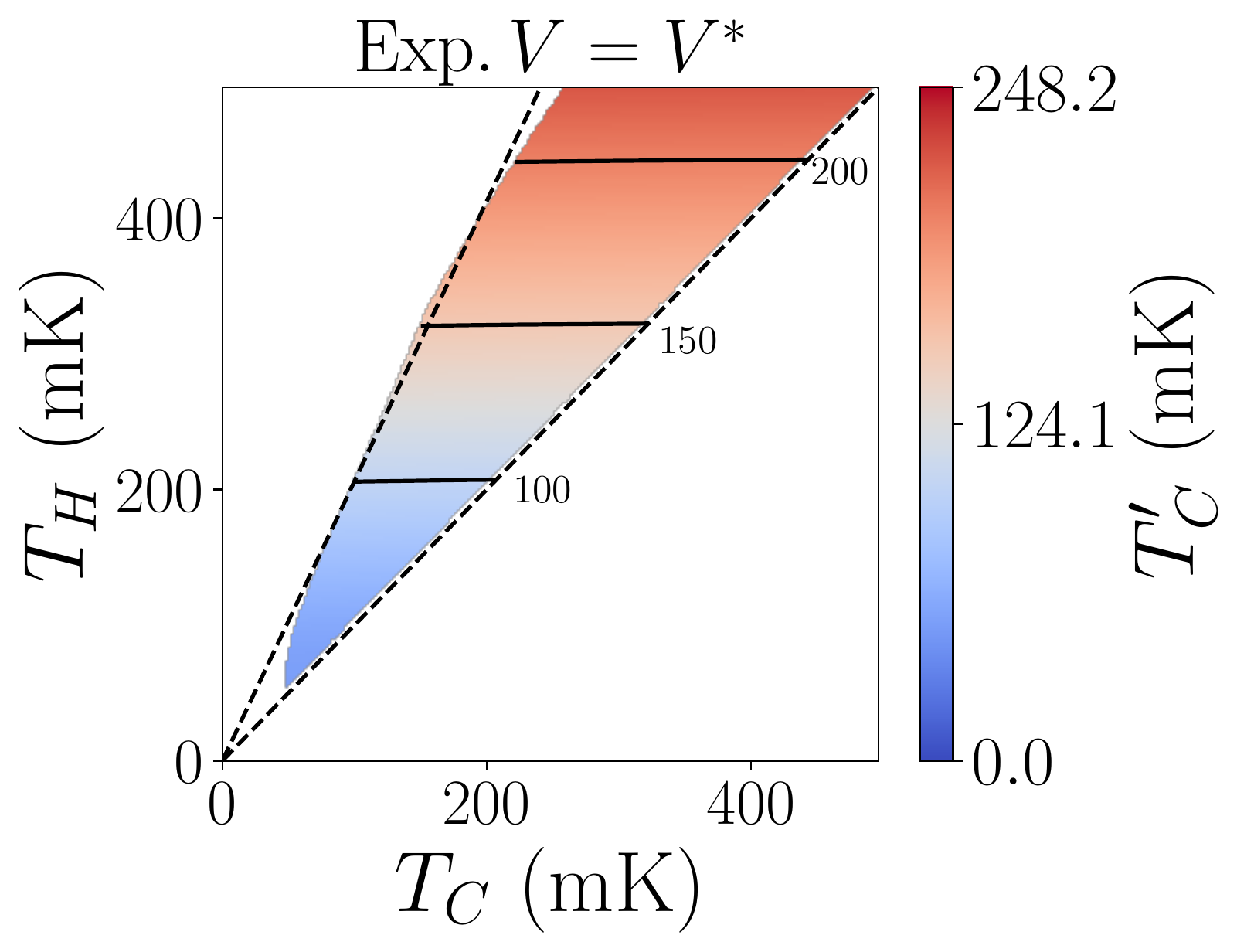}
	\caption{Final temperature $T'_C$ of the cold qubit as a function of initial temperatures $T_H,T_C$, in the refrigeration region, for $V=V^*$.}
	\label{fig:figure3}
\end{figure}

Figure \ref{fig:figure3} shows the final temperature $T_C'$ of the cold qubit, as a function of initial temperatures $T_H$ and $T_C$,
in the refrigeration region [R], for the $V=V^*$ case.   

\subsection{Methods}
We implemented the cooling protocol on the IBM Quantum processor \textit{ibmq\_jakarta} which we remotely accessed through the Qiskit library.\cite{Qiskit-Textbook} The topology of the quantum processor is shown in Fig. \ref{fig:figure1}b). Only qubits $q_0$, $q_1$ and $q_2$ were addressed in our experiments.

Four sets of experiments were performed each with the qubits being initialized in one of the states $\ket{00,0}$, $\ket{00,1}$, $\ket{11,0}$ and $\ket{11,1}$ forming the so called-computational basis.  
After initialization, the gate $U$, with either the choice $V=\mathbb{I}$ or $V=V^*$, as described above, was applied. 
For the $V=\mathbb{I}$ case we let the IBM compiler find a decomposition in elementary gates, which was then applied to the  hardware, whereas in the case $V=V^*$ the decomposition in Fig. \ref{fig:figure1}c) was directly applied.
Finally, projective measurement in the computational basis were performed, and the outcome recorded. This procedure was repeated 
$\mathcal{N}=8192$ times for each initial state and choice of $V$ to obtain the statistics $p_{i'j',k'|ij,k}$ that the compound system ends up in state $\ket{i'j',k'}$ given that it was prepared in state $\ket{ij,k}$. These data were error mitigated following a calibration performed before the experiments, accordingly to the standard procedure described in Refs. \citep{Qiskit-Textbook,santini2021experimental}.

The energy variations of the two subsystems and the total work were computed as
\begin{subequations}
	\begin{align}\label{eq:deltaEs}
		\me{\Delta E_H} &= \sum_{\mathbf{i},\mathbf{i}'} \widetilde{E}^H_{\mathbf{i}'} p_{\mathbf{i}'|\mathbf{i}} p_\mathbf{i} - \sum_\mathbf{i} \widetilde{E}^H_\mathbf{i} p_\mathbf{i},\\
		\me{\Delta E_C} &= \sum_{\mathbf{i},\mathbf{i}'} E^C_{\mathbf{i}'} p_{\mathbf{i}'|\mathbf{i}} p_\mathbf{i} - \sum_\mathbf{i} E^C_\mathbf{i} p_\mathbf{i},\\
		\me{W} &=  \me{\Delta E_H} + \me{\Delta E_C},
	\end{align}
\end{subequations}
where $\mathbf{i}$($\mathbf{i}'$) is a short notation for the multi index set $i,j,k(i',j',k')$. The symbols $\widetilde{E}^{H}_\mathbf{i}$, $E^C_\mathbf{i}$ denote, respectively the hot and cold subsystem eigenenergies, reading, for the cold subsystem $E^{C}_{ij0}=- \hbar\omega_C/2,E^{C}_{ij1}= \hbar\omega_C/2$ and, for the hot subsystem  $\widetilde{E}^H_{00k}=-\hbar\omega_H/2,\widetilde{E}^H_{11k}=\hbar\omega_H/2, \widetilde{E}^H_{10k}=\Delta/2, \widetilde{E}^H_{01k}=-\Delta/2$, where $\Delta = \hbar\omega_{0}-\hbar\omega_{2} \approx 0.13\ \mathrm{GHz}$ is the detuning between between qubit $0$ and qubit $2$. Note that the actual hot subsystem Hamiltonian $\widetilde H_H= \sum_{\mathbf{i}} \widetilde{E}^H_\mathbf{i} \ket{\mathbf{i}}\bra{\mathbf{i}}$, differs from the ideal Hamiltonian $H_H$, Eq. (\ref{eq:HH}). For the initial distribution $p_\mathbf{i}$ we used the expression
\begin{align}
p_\mathbf{i}=e^{-\beta_H E^H_\mathbf{i} -\beta_C E^C_\mathbf{i}}/(Z_H Z_C)
\label{eq:p-i}
\end{align}
with $E^H_\mathbf{i}$ the eigenvalues of the ideal Hamiltonian, Eq. (\ref{eq:HH}). 
We remark that this procedure amounts to create the initial bi-thermal preparation artificially, rather than physically, a method that is often used in quantum thermodynamics experiments, see e.g.,\cite{Hernandez21NJP23}.

The plots in Fig. \ref{fig:figure2} were obtained by looking at the signs of $\av{\Delta E_H},\av{\Delta E_C},\av{W}$, for both theory and experiment. The region [H] is the region  $\av{\Delta E_H}>0,\av{\Delta E_C}>0$; the region [E] is the region  $\av{W}<0$; the region [A] is the region $\av{\Delta E_H}<0,\av{\Delta E_C}>0,\av{W}>0$; the region [R] is the region  $\av{\Delta E_C}<0$.

The final temperature of the cold qubit, reported in Fig. \ref{fig:figure3}, was calculated according to the formula
\begin{align}
k T'_C = - \hbar\omega_C [\ln(Q/(1-Q))]^{-1}
\end{align}
with $Q$ being the final population of state $\ket{1}_C$ of the cold qubit, namely: $Q= \sum_{ij} p'_{ij1}$, where $p'_\mathbf{i}=\sum_\mathbf{i} p_{\mathbf{i}'|\mathbf{i}} p_\mathbf{i}$.

\section{Purification method}
In its essence, the above described method is the two qubit SWAP engine with the only difference that the role of the hot qubit is played by the ground and most excited state of a two-qubit compound. We remark that for the two qubit SWAP engine, in the refrigeration regime, $\beta_C \omega_C < \beta_H \omega_H$, the hot qubit initially has a higher degree of purity than the cold qubit. To see that, let $p_{r}^g=e^{\beta_r \omega_r/2}/[2 \cosh(\beta_r \omega_r/2)]$ be the probability to initially find the qubit $r=H,C$ in its excited state. Note that the function $f(x)=e^{x}/[2\cosh(x)]= 1/(e^{-2x}+1)$ is monotonously increasing, hence the condition $\beta_C \omega_C < \beta_H \omega_H$ implies 
\begin{align}
p_H^g > p_C^g\, ,
\end{align}
meaning that, despite being hotter, the qubit $H$ is in fact initially purer than the qubit $C$. Swapping the populations, then results in cooling the latter. On the basis of this observation, one might object that if you have a qubit that is initial purer than another, then, from a practical point of view, the best option would be to disregard the less pure qubit and use the purer one in your quantum circuit: applying an operation that swaps its population with that of another qubit can only degrade the initially available purity.

A further observation is that in our practical implementation where the hot qubit is replaced by a compound of two qubits, we have assumed, as detailed above, that only its ground and most excited states are populated. That is an idealisation that does not realistically adhere to what would happen in practice. In reality, those intermediate states exist and have some finite population.

In sum, the above described method is too idealised and does not allow to improve qubit purity.
These practical considerations lead us to the idea of considering the more realistic case where all physical qubits participating in the cooling procedure have some finite temperature. 

It is not hard to see that with three identical qubits all prepared at the same temperature, hence with same degree of purity, application of the unitary gate $U$ introduced above, with $V=\mathbb{I}$, will result in qubit $q_1$ being cooled (i.e., getting purified), at the cost of heating up the other two: this is a possibility that only a 3 (or more) qubit design offers. 
To see that, let $x$ denote the probability that any qubit initially is in its ground state. The effect of the unitary gate in Eq. \ref{eq:U} with $V = \mathbb{I}$ is to swap the population of the states $|11,0\rangle$ and $|00,1\rangle$. Using the notation $p'_{ij,k}$ to denote the post gate probabilities, we have
\begin{align}
		p'_{00,1}=p_{11,0} = (1-x)^2 x ,\, p'_{11,0}= p_{00,1} = (1-x)x^2.
\end{align}
Consequently the ground state population of qubit $q_1$ after the unitary evolution reads
\begin{align}
	x' = \sum_{i,j} p'_{ij,0} =  3x^2 - 2x^3.
\end{align}
Note that $x'>x$ for $x>1/2$, and $x'<x$ for $x<1/2$. This means that the protocol in fact enhances the purity of qubit $q_1$, already in the case of identical temperatures.

\subsection{Results}

\begin{figure}[t]
	\includegraphics[width=\linewidth]{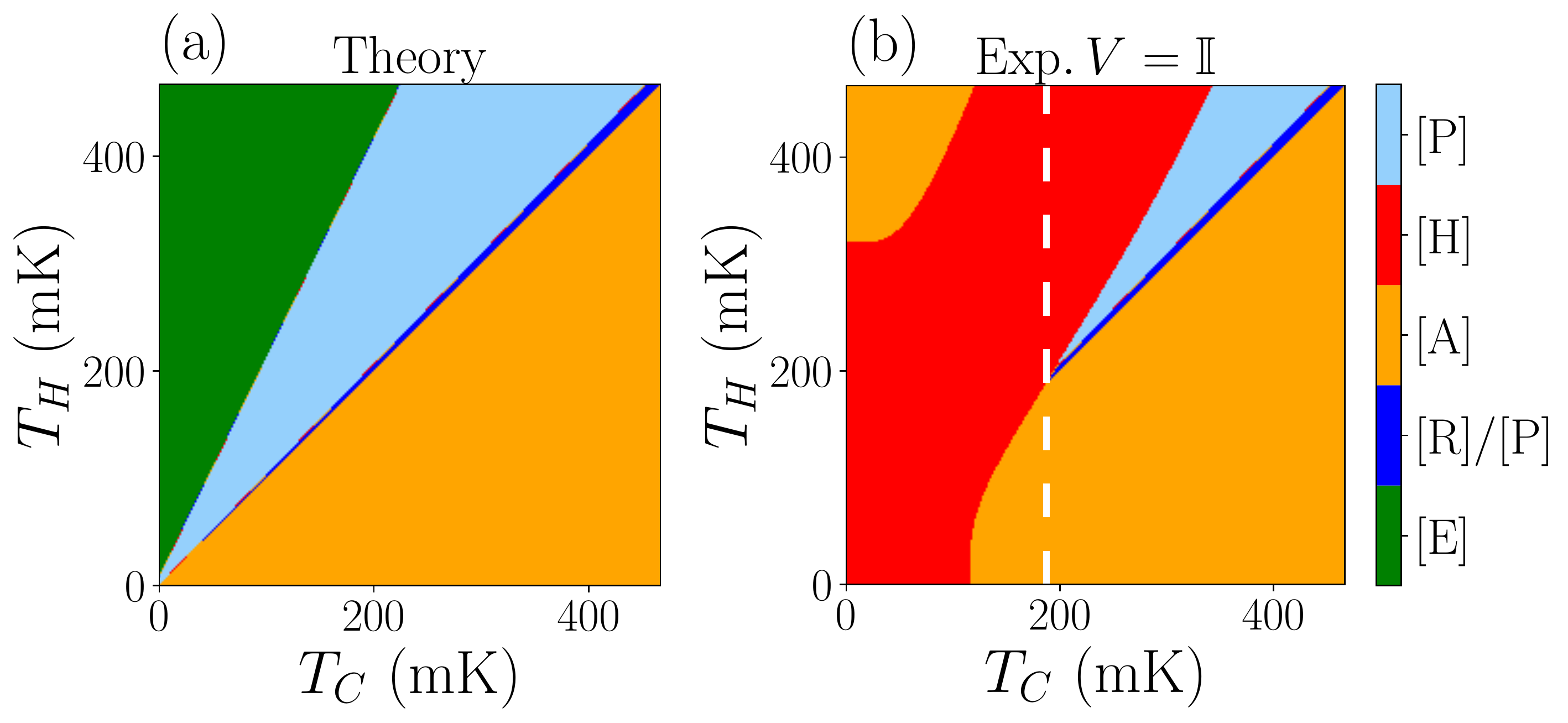}
	\caption{``Phase diagrams'' of the quantum heat engine implementing our purification method. Panel a): Theory. Panel b): Experiment. The union of blue and light blue regions is the refrigeration region [R]. The light blue region [P] is the purification region.}
	\label{fig:figure4}
\end{figure}

We apply this idea to three qubits of the IBM \textit{ibmq\_casablanca} QPU, whose topology is as the one depicted in Fig. \ref{fig:figure1}b), featuring almost identical resonant frequencies, namely $\omega_0 \approx 4.82 \mathrm{GHz}$, $\omega_1 \approx 4.76 \mathrm{GHz}$ and $\omega_2 \approx 4.90 \mathrm{GHz}$. 
Specifically, qubits $q_0,q_2$ where prepared a temperature $T_H$, and qubit $q_1$, the one we want to cool, was prepared at temperature $T_C$ (see Fig. \ref{fig:figure1}).
Figure \ref{fig:figure4}a) shows the theoretical phase diagram, while the figure \ref{fig:figure4}b) shows the result of our experiments.
Note that in both cases a light blue region which we dub ``purifier'' [P] appears. That is the subset of the [R] region where qubit $q_1$ was not only refrigerated, but also ended up in a state of higher purity than each qubit initially had. Note also the blue strip around the line $T_C=T_H$ signaling the possibility of cooling a qubit using more qubits at the same temperature. The region [P] did not extend to the $T_C=T_H$ line in Fig. \ref{fig:figure4}) due to the slight mismatch of qubits resonant frequencies, see details below.

As with the original method, gate noise is the main obstacle towards effective implementation of the method: note the presence of the [H] region in the experimental data, which is not present in the theoretical phase diagram. The main difference with the previous method is that now the intermediate states $\{
\ket{01,0}
\ket{01,1}
\ket{10,0}
\ket{10,1}
\}$
are populated. Despite that hinders the freedom of choosing $V$ (hence of minimising gate noise), that is in fact a realistic condition and also unlocks the, otherwise excluded, possibility of purification.

\begin{figure}[t]
	\includegraphics[width=\linewidth]{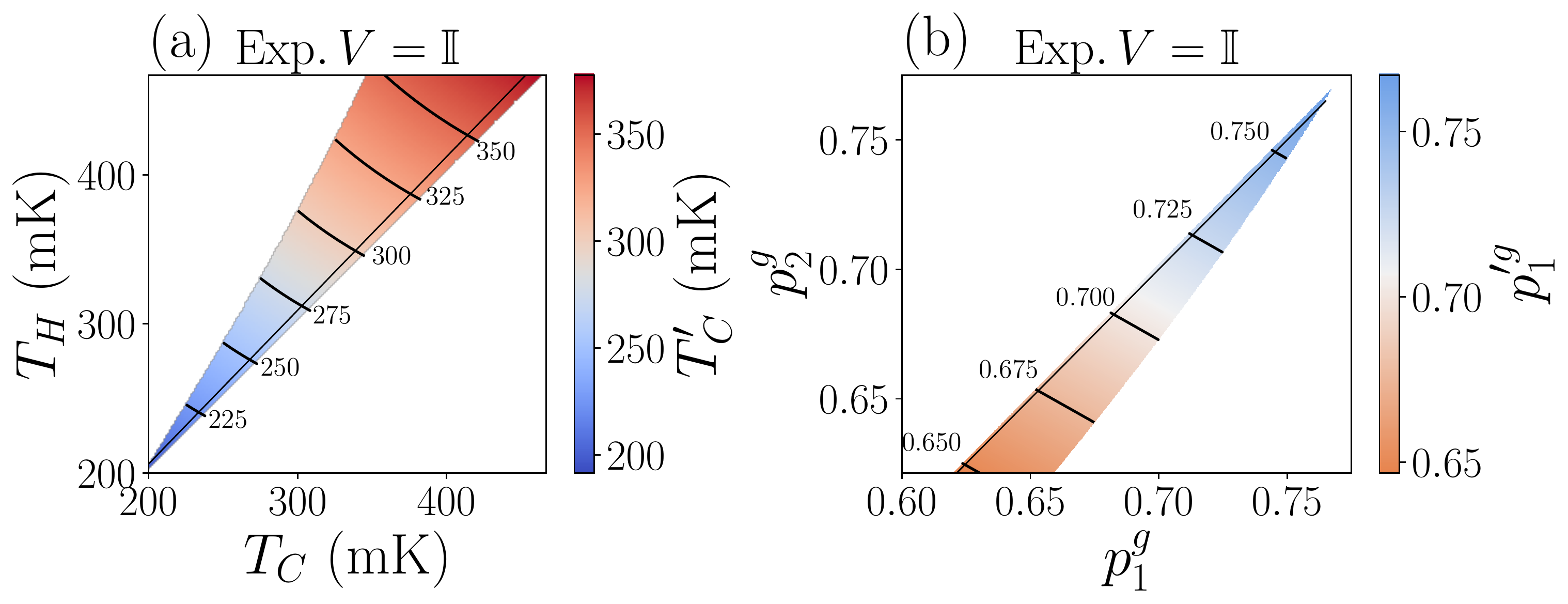}
	\caption{Panel a): Final temperature of qubit $q_1$ as a function of its initial temperature $T_C$ and the common initial temperature, $T_H$ of qubits $q_0,q_2$ in the [R] region. The purification region [P] is above the straight thin black line.
	Panel b) Final ground state population $p'^g_1$, of target qubit $1$ as a function of its initial population $p^g_1$, and the population $p^g_{2}$ of $q_2$ (i.e., the purest of the equal temperature hot qubits). The purification region [P] is below the straight thin black line.
	}
	\label{fig:figure5}
\end{figure}

Figure \ref{fig:figure5}a) depicts the final temperature of qubit $q_1$ as a function of its initial temperature $T_C$ and the common initial temperature, $T_H$ of qubits $q_0,q_2$ in the [R] region. Figure \ref{fig:figure5}b) depicts, in the same region [R], the final ground state population $p'^g_1$, as a function of its initial population $p^g_1$, and the population $p^g_{2}$ of $q_2$. Like in Fig. \ref{fig:figure5}a) qubit $q_0$ was prepared at the same temperature of $q_2$, hence, its ground state probability $p^g_0$ is in one to one correspondence with $p^g_2$, and given that $\omega_2 > \omega_0$, it is $p^g_{2}>p^g_{0}$, so $q_2$ was initially purer than $q_0$. Note the extended region, below the straight black line, where it is $p'^g_{1}>p^g_{1}>p^g_{2}>p^g_0$. As detailed below that is the region [P] where we can claim that the target qubit $q_1$ ended up at a higher purity than any qubit initially had.

\subsection{Methods}
The method was implemented on the \textit{ibmq\_casablanca} QPU, whose topology (which is identical to that of the \textit{ibmq\_jakarta} QPU) is depicted in Fig. \ref{fig:figure1}b). Differently from the previous method, now eight sets of experiments were performed each with the qubits being initialized in one of the states of the complete three qubits energy eigen-basis $\{\ket{ij,k}\}$, with $i,j,k=0,1$. The results were error mitigated in the same manner. The initial state $\rho$ was prepared artificially by weighting 
the preparation states $\ket{\mathbf{i}}$ according to the probabilities
\begin{align}
p_\mathbf{i}=e^{-\beta_H E^{02}_\mathbf{i} -\beta_C E^1_\mathbf{i}}/(Z_H Z_C)
\label{eq:p-i}
\end{align}
where $E^{02}_\mathbf{i}$, $E^1_\mathbf{i}$ are the eigenvalues of the hot and cold systems Hamiltonians reading
\begin{align}
	H^{02} = &-\frac{\hbar\omega_0}{2}(|0\rangle\langle 0|_0-|1\rangle\langle 1|_0)\otimes\mathbb{I}_{2}\otimes\mathbb{I}_{1} \notag
	\\
	&-\mathbb{I}_{0}\otimes\frac{\hbar\omega_2}{2}(|0\rangle\langle 0|_2-|1\rangle\langle 1|_2)\otimes\mathbb{I}_{1}
	\\
	H^1 = &-\mathbb{I}_{0}\otimes\mathbb{I}_{2}\otimes\frac{\hbar\omega_0}{2}(|0\rangle\langle 0|_1-|1\rangle\langle 1|_1).
\end{align}
where $Z_{02} = \Tr e^{-\beta_H H^{02}} , Z_{1} = \Tr e^{-\beta_C H^{1}}$ are the according canonical partition functions.

The unitary $U$ with $V=\mathbb{I}$ was implemented and applied as described for the previous method, so as to give the transition probabilities $p_{\mathbf{i}'|\mathbf{i}}$, which were used to calculate the energy changes of the hot and cold subsystem as in Eq. (\ref{eq:deltaEs}), but now with the eigenvalues $E^{02}_\mathbf{i}$ for the hot system, and $E^1_\mathbf{i}$ for the cold part.
These were used to produce the plot in Fig. \ref{fig:figure4}b), according to the rules defined above. The [P] region is the subset of the [R] region where the purity of qubit $q_1$ increased beyond the initial purity of all the available qubits.

Fig. \ref{fig:figure4}a) was produced using the theoretical values (zero gate noise) of the energy changes, reading:
\begin{align}
	\langle\Delta E_H\rangle =& \frac{\Omega}{4}f(\beta_H\Omega,\beta_C\omega_1)g(\beta_H\omega_0,\beta_H\omega_2)	\\
	\langle\Delta E_C\rangle= &-\frac{\omega_1}{4}f(\beta_H\Omega,\beta_C\omega_1)g(\beta_H\omega_0,\beta_H\omega_2)
\end{align}
where $f(x,y)=\tanh(x/2)-\tanh(y/2)$, $g(x,y)=1+\tanh(x/2)\tanh(y/2)$,
and $\Omega =\omega_{0}+\omega_{2}$. These lead to the following analytical expressions for the various operation regions displayed in Fig. \ref{fig:figure4}a):
\begin{align}
	&[E] \quad T_H\geq(\Omega/\omega_{1})T_C;\\
	&[R] \quad T_C\leq T_H\leq(\Omega/\omega_{1})T_C;\\
	&[A] \quad T_H\leq T_C;\\
	&[P] \quad T_C\max\{\omega_{0}/\omega_{1},\omega_{2}/\omega_{1},1\}\leq T_H\leq(\Omega/\omega_{1}) T_C.
\end{align}
Note that if the three qubits have identical level spacings $\omega_{0} = \omega_{1} = \omega_{2}$, than the $[P]$ region coincides with the $[R]$ region.

It is important to remark that the purity of a quantum state is defined as $P[\rho]= \Tr \rho^2$. In our experiments we have accessed the projection of qubit $q_1$ final state, call it $\rho'$, on its computational basis, that is we addressed the state $\Pi[\rho']= p'^g_1 \ket{0}_1\bra{0}+(1-p'^g_1)\ket{1}_1\bra{1}$, which might not coincide with $\rho'$. Its purity reads $P[ \Pi[\rho']]=2(p'^{g}_{1})^2-2p'^g_1+1$ 
and it is larger than the purity of the initial diagonal state $\rho=p^g_1 \ket{0}_1\bra{0}+(1-p^g_1)\ket{1}_1\bra{1}$, if the final ground state population $p'^g_1$ is larger than  the initial population $p^g_1 > 1/2$.
The light blue region [P] in Fig. \ref{fig:figure4}b) is the region where that happens, namely, to be precise, it is the region where the purity of the projected state of qubit $q_1$ increased, namely $P[\Pi[\rho']]\geq P[\rho]$. However it can be proved that $P[\rho'] \geq P[\Pi[\rho']]$,\footnote{To prove that for a generic density matrix $\sigma$ it is $P[\sigma] \geq P[\Pi[\sigma]]$, note that $P[\sigma]= e^{-S_2[\sigma]}$, where $S_2[\sigma]$ is the order 2 quantum Renyi entropy, and that the latter obeys the data processing inequality $S_2[\mathcal N[\sigma]]\geq S_2[\sigma]$, with $\mathcal N$ a unital channel. \cite{Chehade19SCLP14} The inequality $P[\sigma] \geq P[\Pi[\sigma]]$ then  follows from noting that the projection $\Pi$ is a unital channel and that the function $e^{-x}$ is monotonically decreasing.} thus in the [P] region it is $P[\rho'] \geq P[\rho] $, which allows us to claim that the purity has increased in the region [P] (it possibly increased in a larger region, though).

\section{Discussion}

We have investigated a thermodynamic method to purify a qubit on a QPU (based on quasi-identical qubits), at the expense of heating up two other qubits. 
Our starting point is the implementation of the two qubit SWAP engine with three qubits. We have shown that the method is limited by gate noise. Our implementation on a QPU, evidencing a cooling capability down to $52$ mK has to be taken with a grain of salt, as it was obtained assuming the unrealistic condition that the intermediate states $\ket{01}_H,\ket{1,0}_H$ of the hot subsystem were not pupulated, which is a rather drastic idealisation. 
Considering the realistic scenario where all three physical qubits are prepared in thermal states, and the intermediate states are accordingly populated, unlocks enhanced refrigeration possibilities. In particular, with the specific choice of cooling gate with $V=\mathbb{I}$, it is possible to genuinely increase the purity of a chosen qubit beyond the initial level of purity of all qubits participating in the process. Our implementation on a real QPU evidenced a purification capability down to about $200$ mK.
This value could be further decreased with less noise on the gate: theoretically, with zero noise on the gate, purification would be possible at any temperature.
We remark that on the QPU that we used  the actual physical temperature of qubit initialisation was about $15$ mK, which is way below the limit of $200$ mK observed in our implementation. 

Summing up, our results represent a first step towards the development of practical thermodynamic methods to purify qubits on QPU's. In particular, they evidence the necessity to employ at least two extra qubits to purify one qubit, and single out gate noise as the main obstacle on the way to practical application.

\subsection*{Acknowledgments}
We acknowledge the use of IBM Quantum services for this work.\cite{IBMQ_ref} The views expressed are those of the authors, and do not reflect the official policy or position of IBM or the IBM Quantum team. In this paper we used \emph{ibmq\_jakarta} and \emph{ibmq\_casablanca}, which are IBM Quantum Falcon r5.11H Processors. 
Andrea Solfanelli and Alessandro Santini acknowledge that their research has been conducted within the framework of the Trieste Institute for Theoretical Quantum Technologies (TQT). The authors wish to thank Prof. C. Jarzynski for insightful comments on the manuscript.

\section*{Author declarations}
\subsection*{Conflicts of interest}
The authors have no conflicts to disclose.

\section*{DATA AVAILABILITY}
The data that support the findings of this study are available from the corresponding author upon reasonable request.

\end{document}